\documentclass[article,12pt,onecolumn,showpacs,preprintnumbers,amsmath,assume,plain]{revtex4}
\usepackage{graphicx}
\usepackage{dcolumn}
\usepackage{bm}
\usepackage{verbatim}
\usepackage{hyperref}
\usepackage{amsmath}
\usepackage{float}

\def\bG{ {\bf $\Gamma^+$} }
\def\bw{ {\bf $\tilde{\omega}$} }
\def\bg{ {\bf $\gamma$} }

\def\re{ \,{\rm Re}\, }

\begin{document}

\title{A non-linear minimization calculation of the renormalized frequency\bw in dirty d-wave superconductors}

\author{P. Contreras $^{1}$}
\affiliation{$^{1}$ Departamento de F\'{\i}sica and Centro de F\'{\i}sica Fundamental, Universidad de Los Andes, M\'erida 5101, Venezuela}
\author{Juan Moreno $^{2}$}
\affiliation{ $^{2}$ Facultad de Ingenier\'{\i}a, Universidad de la Empresa, Montevideo 11300, Uruguay}
\date{\today}

\begin{abstract}
This work performs a comparative numerical study of the impurity average self-frequency \bw in an unconventional superconducting alloy
with non-magnetic impurities. Two methods are used: the Levenberg-Marquardt algorithm as a non-linear minimization problem,
and a fixed-point iteration procedure. The unconventional superconducting renormalized by impurities \bw is a self-consistent complex non-linear equation with two varying parameters: the impurity concentration \bG and the strength of the impurities c, for which its numerical solution is a computational challenge.
This study uses an order parameter that corresponds to the high-temperature superconducting ceramics (HTS) with a well-established gap symmetry.
The results reveal the computational efficiency of the non-linear minimization technique
by improving the calculations of the \bw computation when using a 2D parameter space (\bG, c),
particularly in the unitary regime, where the imaginary part of \bw is a complicated expression of those parameters;
this allows to enhance the study of the universal behavior of this particular quantum mechanical state.

\end{abstract}

\maketitle
\newpage
\section{Introduction}\label{sec:intro}

The effect caused by non-magnetic impurities in unconventional superconductors play a fundamental role
in the understanding of their physical properties when these materials are doped.
We use an order parameter well established for (HTS) \protect\cite{1}.
Superconducting ceramics have a transition temperature
T$_c$ close to the boiling point of liquid Nitrogen T$_{Ni}$ = -195.79 degrees Celsius \protect\cite{2}.
Although these materials are fragile in their elastic properties, they aim to revolutionize
the technology of electrical conduction without energy loss \protect\cite{3}.
In particular, these materials lose their superconduction properties as non-magnetic impurities are added.
For example for $YBCO$ ceramics doped with $Zn$ impurities, the transition temperature T$_c$ begins to decrease rapidly \protect\cite{1}.

\begin{figure}[ht]
\includegraphics[width = 2.5 in, height= 2.5 in]{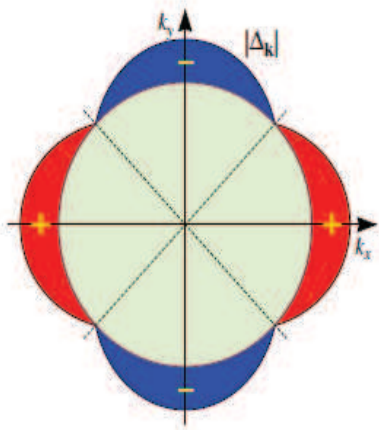}
\caption{\label{1} $d_{x_2-y_2}$ gap for the one dimensional irreducible representation $B_{1g}$
of the tetragonal point symmetry group $D_{4h}$ for a spherical Fermi surface.}
\end{figure}

We know that for (HTS) the energy gap corresponds to a paired singlet quantum state with a
$d_{x_2-y_2}$ symmetry, for example in hole-doped cuprate superconductors such as YBa$_2$Cu$_3$O$_7$ and
Tl$_2$Ba$_2$CuO$_6$ \protect\cite{4,8}.
The superconducting gap for this symmetry has lines nodes on the Fermi surface, and the energy gap corresponds to the one-dimensional
irreducible representation $B_{1g}$ of the tetragonal point symmetry group $D_{4h}$.
They are called $d$-wave superconductors and they have elastic impurity scattering preserving the total kinetic energy.
The d-wave gap changes sign as a function of the azimuthal angle $\theta$
with line nodes in the energy spectra as it is illustrated in Fig.~\ref{1}.

The task of this work is to obtain the solution of the equation for \bw for a wide range of (\bG, c) parameters.
This complex non-linear two-dimensional equation is very difficult to solve by conventional iterative algorithms.
Instead of that, in this work we solve \bw using two numerical routines; throughout this study \bw is the renormalized frequency,
\bg represents the inverse of the residual average lifetime $\tau$ at zero frequency, and $N/N_0$ is the normalized superconducting
density of states (DOS). We calculate and studied the DOS at low energies.
The DOS gives insight information about the temperature behavior of some relevant
thermodynamic and kinetic quantities in unconventional
superconductors that are experimentally measured such as the specific electronic heat $C(T)$,
the thermal conductivity $\kappa_{ij} (T)$ and the sound attenuation $\alpha_{ij} (T)$ \protect\cite{p0,p1,p2}.
These measurements help to clear up the structure of the gap symmetry.

This work consists of five sections. The first section introduces the subject of the dirty d-wave (HTS) and limits our study to an
spherical (isotropic)
Fermi surface with a line-nodes order-parameter of the one-dimensional representation $B_{1g}$ of the tetragonal group $D_{4h}$.
The second section explains shortly the \bw formalism given by \textbf{Mineev and Samokhin} \protect\cite{5}, and
also follows the \textbf{Carbote and Schachinger} \protect\cite{12} approach which do some of these numerical calculations in a different
physical context (the residual absorption at zero temperature of the optical spectral weight).
Briefly, we derive and explain equation (2) and equation (4) and refers the lectors to these two references for further details.
The third section briefly introduces and explains the way the two algorithms work and
their main differences (the implementation advantages and$/$or disadvantages).
The fourth section presents the numerical solutions of equation (2) and (4) by comparing the
fixed-point iteration procedure and the non-linear minimization algorithm.
The fifth section performs a numerical evaluation of the normalized $DOS$ in the unitary and Born regimes.
The calculation of the low energy DOS is crucial for the impurity case, because it shows the limit which
belongs to the (HTS) d-wave superconductors. We conclude with a summary of the main results and some final thoughts
are provided for further research.


\section{\bw formalism for non-magnetic impurities}\label{sec:nmi}

In the case of unconventional superconductors with a singlet pairing state such as (HTS),
when non-magnetic impurities effects are added, the interaction Hamiltonian between
impurities and Cooper pairs is given by the following expression:

\begin{equation}
    \label{11}
    H = U_0 \sum_{i,j} \hat{c}^{+}_{f} \hat{c}_{i},
\end{equation}
where U$_0$ is the interaction potential, $\hat{c}_{i}$ and $\hat{c}^{+}_{f}$ are the annihilation and creation operators of the cooper pairs.
When multiple elastic scattering of a Cooper pair occurs \protect\cite{5,walker,19};
\bw is not only given in the Born limit (weak potential with $U_0 \ll 1$ and a Cooper pair scattered by one impurity).
Instead, a Cooper pair scatters many impurities (strong Coulomb interaction with $U_0 \gg 1$) and the calculation of the equation
for \bw has to be done in the unitary regime. However, the fixed-point method has problems finding a proper \bw.

We consider the important case of weak disorder (restricting our calculation more). It means that at sufficiently low concentration
of impurities, the condition $k_f l \gg 1$ where $l$ the mean free path prevails and crossed Feynman diagrams are ignored.
Physically, this means that there is no interference between scattered waves having different probability amplitudes. 
\cite{5,9,10,11,19}.

We further consider the dimensionless Planck units ($\hbar = c = k_B = 1$) for the whole calculation.
Nerveless, the renormalized frequency \bw does not depend on angle,
but because of the elastic impurity scattering changes (elastic scattering is the only scattering process accounted
for in our work). \textbf{The dimensionless self-consistent frequency equation for} \bw in a t-matrix approximation is \protect\cite{5,12}

\begin{equation}
  \label{22}
  \tilde{\omega}(\omega) = \omega + i \pi \Gamma^+ \; \frac{\left\langle g(\theta,\tilde{\omega}) \right\rangle_{FS}}
  {c^2 + \left\langle g(\theta,\tilde{\omega}) \right\rangle^2_{FS}}.
\end{equation}

Where in equation (2) the imaginary term is the self-energy renormalized for non-magnetic doped impurities
$\Sigma_{imp} (\tilde{\omega})$ and the Fermi averaged expression for $g (\tilde{\omega})$ is

\begin{equation}
    \label{33}
        g(\tilde{\omega}) = \Big{\langle} \frac{\tilde{\omega}}{\tilde{\omega}^{2} - |\Delta|^2} \Big{\rangle}_{FS},
\end{equation}
in equation (2) the parameter $c = 1 / (\pi \; N_F \; U_0)$ and the parameter  $\Gamma^+ = n_{imp} / (\pi \; N_F)$
For very low frequencies $(\omega \rightarrow  0)$, $\tilde{\omega} = i \gamma$ and $\gamma$ becomes equal to
\begin{equation}
  \label{44}
  \gamma(\Gamma^+,c) = \pi \; \Gamma^+ \frac{g(i \; \gamma)}{c^2 + g^{2}(i \;\gamma)},
\end{equation}
where
\[
g(i \; \gamma) = \Big{\langle} \frac{\gamma}{\gamma^{2} + |\Delta|^2} \Big{\rangle}_{FS}.
\]

We call equation (4) \textbf{the transcendental equation of the residual average lifetime at zero frequency}
with $\gamma = 1 / \tau (0)$.
According to equation (2), we know that \bg depends on \bG and c,
so first we solve and analyze this dependency numerically for equation (4).

We write the imaginary part of equation (2) as $Im [\tilde{\omega} (\omega)] = 1 / \tau (\omega)$,
where $\tau (\omega)$ represents the average lifetime for the bounded quasi-stationary state
and $Im [\tilde{\omega} (\omega)]$ is the disintegration probability $W$ per unit time \protect\cite{19}.
In other words, the calculation of the imaginary term empowers a direct evaluation
of the average lifetime in a particular quasi-stationary state for a weak-elastic scattered and doped
d-wave HTC superconductor.

\section{Algorithms}\label{sec:nm}

We used two different algorithms to find the numerical solution of equation (2):
An iterative fixed-point method with error differences to find the sought solution for \bw \cite{16}, and a
minimization routine based on the non-linear Levenberg-Marquardt method \protect\cite{14,17}.
Both were developed in a C language (Standard C11) \cite{15,16}, which supports complex numbers in a native way, in
the $GCC$ compiler (GNU Compiler Collection) and the open-source integrated development environment NetBeans \protect\cite{18}.


\subsection{Fixed-point method}
The \textbf{fixed-point algorithm} \cite{16} iterates a maximum number of times \textbf{max\underline{ }iter} equation (2).
It evaluates whether the real and complex parts of equation (2) are smaller than a specified tolerance \textbf{TOL} between two continuous evaluations of the equation. When the smaller values are found, it converges and the solution is reached.
Its performance and solution is strongly associated with the "quality" of the initial conditions and the dimensions of the parameter space
involved (c,\bG) as we conclude in this work.

\subsection{Levenberg-Marquardt algorithm}
The \textbf{Levenberg-Marquardt} algorithm \protect\cite{14,17} is an iterative method that solves
non-linear quadratic systems through the combination of the descending gradient and the Gauss-Newton methods
by following the behavior of the quadratic error.
The method provides a solution for the minimization of linear complex quadratic systems of equations.
This implies that the minimization function must have the following special form: \[f(x) = \frac{1}{2} \sum_{j=1}^{m} r^2(x) \],
where $x = (x_1,...,x_m)$ is a vector, and each $r_j$ is a function $r_j : R_n \rightarrow Rn$.
$r_j$ is known as a residual (it is assumed that $m \geq n$). $f (x)$ in this methodology is represented as a
residual vector $\textbf{r}$ such that $\textbf{r}: R_n \rightarrow R_n$ where $r(x) = (r_1(x), ...,r_m(x))$.
The derivatives of $f(x)$ are written using a Jacobian matrix $\textbf{J}$. The methodology
considers the case where every function $r_j$ is linear, the Jacobian is constant, and $\textbf{r}$
represented as an hyperplane (a 2D space parameter in our case).
$\nabla^2 f(x) = \textbf{J} \textbf{J}^T$ is given by its quadrature, and it is obtained by solving the minimum
when $\nabla f(x) = 0$, in such a case it is found that $x_{min}= -(\textbf{J}^T \textbf{J})^{-1} \textbf{J}^T \textbf{r}$
which is the solution for the normalized equation (2).

Returning to the non-linear case, we have that the distinctive feature of the least-squares problem
is that given the Jacobian matrix $\textbf{J}$, the Hessian $ \nabla^2 f(x)$ can be found.
If this is possible, $r_j$ are approximated by linear functions where the $r_j(x)$ and the $r_j^2(x)$ are small
and the Hessian becomes $ \nabla^2 f(x)= \textbf{J}^T \textbf{J}$  which is the same equation obtained for the linear case.
This common approach applies to systems where the residuals $r_j$ are small (in our case a 2D (\bG, c) space parameter).

\section{Numerical results}\label{sec:nr}

We begin testing the two algorithms to find a solution for equation (2).
Immediately it was discovered that although the
fixed point algorithm has a faster runtime (see \ref{table1}), it does not manage
to differentiate the parameter \bG, obtaining, that no matter the value of \bG is, always the imaginary part of
\bw is very similar for different values of \bG as shown in Fig.~\ref{4}\textbf{(a)}.
This result is explained below, it means that the curve in Fig.~\ref{4}\textbf{(b)}
cannot be reproduced with the fixed-point procedure.

On the other hand, the \textbf{Levenberg-Marquardt} non-linear algorithm
performs a minimization for equation (2) and accurately differentiates small and large values of both parameters:
\bG and c. Thanks to this algorithm, we are able to find the results showed in Fig.~\ref{4}\textbf{(b)}.
In this analysis \bG and c where evaluated for small relative numbers that outline two physical regimes:
the misbehaved ill quasi-particle unitary state \protect\cite{5,12,ult2};
And the Born limit where the elastic scattering, the free mean path, and the quasiparticles states physically are well defined.

Table \ref{table1} summarizes the data processing of both algorithms for a computer architecture
with a 64-bit Intel I5 processor and an 8 GB RAM in a \textbf{Slackware 14.2 OS} \protect\cite{slack}:

\begin{table}[h]
\caption{\label{table1} Runtime for the complex fixed-point and minimization algorithms to calculate $\tilde{\omega}(\omega)$}
\center{\begin{tabular}{|c|c|}
  \hline
  Method & Running performance \\
  \hline
  Self-consistent method& 24 sec\\
  \hline
  Minimization method & 15 min 24 sec \\
  \hline
\end{tabular}
}
\end{table}

\begin{figure}[ht]
\includegraphics[width = 3.5 in, height= 3.0 in]{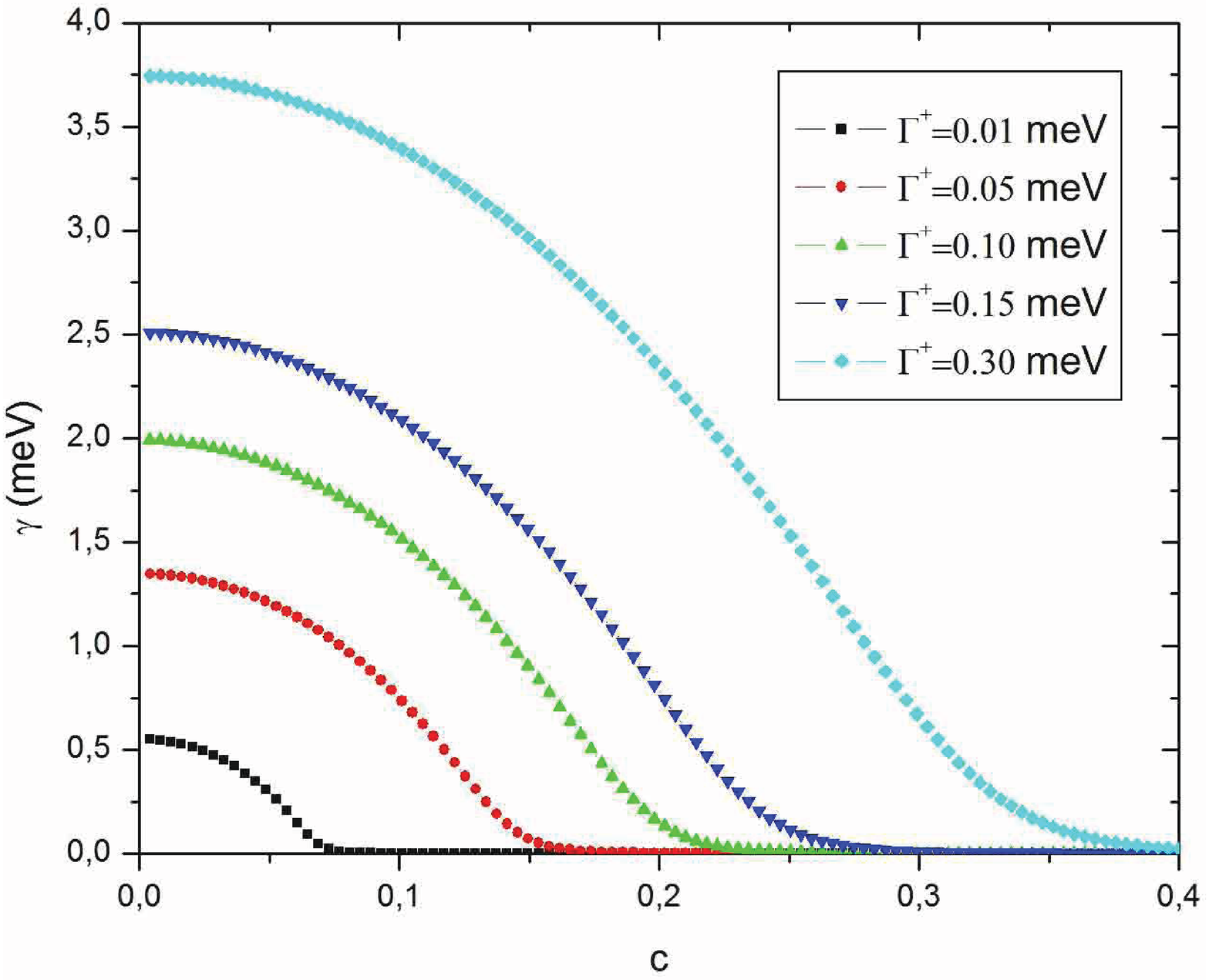}
\caption{\label{2} The transcend equation $\gamma(c)$ as a function of c (strength related parameter) for five
different values of $\Gamma^+$ (concentration related parameter).}
\end{figure}

Subsequently, we analyze the transcendental equation (4) that represents the dispersion of quasiparticles at $\omega= 0$
and for zero temperatures.
From Fig.~\ref{2}, it is observed that when $c = 0$ and U$_0$ rapidly increases, $\gamma$ also increases.
Moreover for \bG $=0.30 meV$ gives the maximum $\gamma$ value in Fig.~\ref{2},
with an observed tendency for decreasing \bG values.
That means that for higher impurity concentration, the Cooper pair lifetime $\tau(\omega) \rightarrow 0$
decreases faster when $c \rightarrow 0$, giving that
for values $c \leq 0.1$ the probability of destroying the superconducting state by the
impurity is calculated correctly.
Nonetheless, for $\gamma$ values with relatively small U$_0$ given when $c \geq 0.1$,
bound states still remain in the sample with a small $\tau(\omega)$.
Consequently, the imaginary part of the zero-frequency dispersion decays faster,
implying that the average lifetime increases faster if the potential strength is not so small.
Fig.~\ref{2} should be compared with reference Fig. 4 bottom frame of \protect\cite{12}
(also Fig.~\ref{2} should be compared with some of the results obtained by Ref.\protect\cite{11} as a reference).

\begin{center}
\begin{figure}[H]
\centering
\includegraphics[width=3.0 in, height=3.0 in]{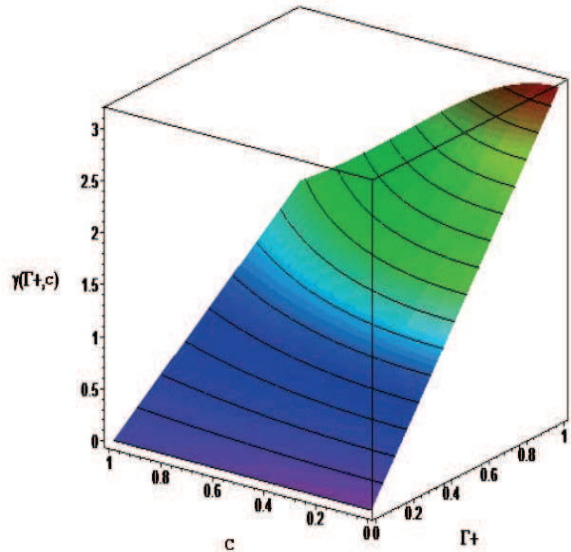}\caption{\label{3} Skecth of the imaginary part
$\gamma(\Gamma^+,c)$ as a continuous 3D function of c and $\Gamma^+$.
The green and red neighborhoods show the smallest lifetimes $\tau(0)$,
and the dark blue part shows the region with the biggest $\tau(0)$ value.
This picture was obtained for gap values closed to the nodes of Fig.~\ref{1}}
\end{figure}
\end{center}

\begin{figure}[ht]
\includegraphics[width = 4.0 in, height= 4.5 in]{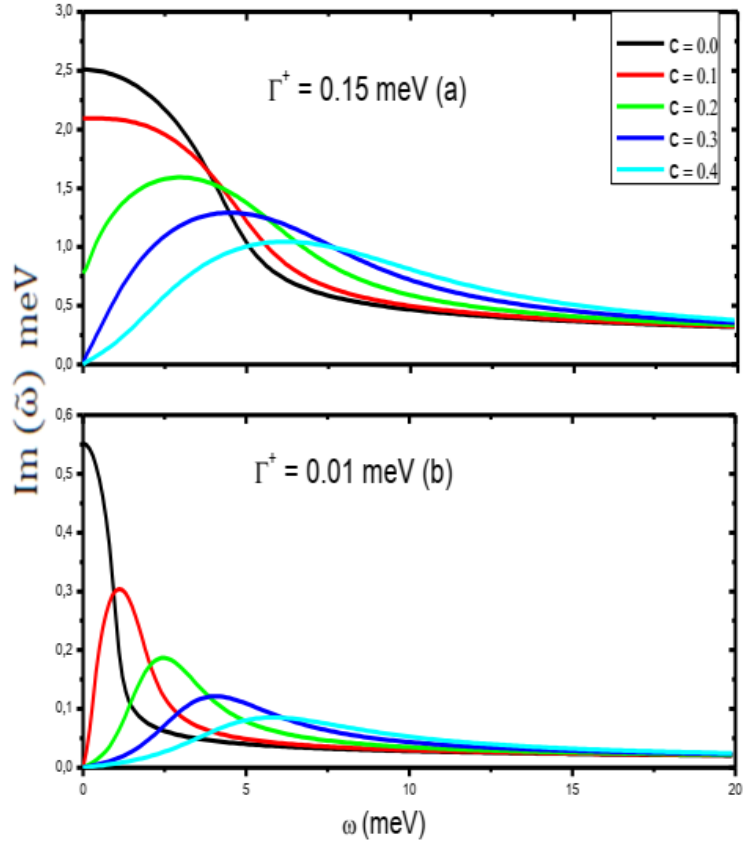}
\caption{\label{4} $Im [\tilde{\omega} (\omega)]$ vs. $\omega$ for small concentrations of impurity
(bottom-frame b, $\Gamma^+=0.01 meV$) and higher concentrations (top-frame a, $\Gamma^+=0.15 meV$)
for the unitary (c=0.0), intermediate(c=0.2) and Born (c=0.4) limits.}
\end{figure}

As a next step, we display the 3D figure of the transcendental equation (4) $\gamma(\Gamma^+,c)$ in Fig.~\ref{3}.
We limit our discussion to places near the minimum gap nodes are located (at zero temperatures),
so in that neighborhood are still ill-defined quasiparticles \footnote{We want to pursue the discussion by Landau and Lifshitz
concerning the resonance spectrum of quasiparticles at quasi-discrete levels by quoting their own words \protect\cite{19}:
\textbf{It may happen, however, that the disintegration probability $W$ of the system
is very small. The simplest example of this kind is given by a particle surrounded
by a fairly high and wide potential barrier.
For such systems with a small disintegration probability, we can introduce
the concept of quasi-stationary states, in which the particles move "inside
the system" for a considerable period of time, leaving it only when a fairly
long time interval $\tau$ has elapsed; $\tau$ may be called the lifetime of the almost
stationary state concerned ($\tau \sim l/W$, where $W$ is the disintegration probability
per unit time). The energy spectrum of these states will be quasi-discrete;
it consists of a series of broadened levels, whose "width" is related
to the lifetime by $\tau \sim \hbar/\Gamma$. The widths of the quasi-discrete
levels are small compared with the distances between them.}}.

It can be seen from Fig.~\ref{3} that a zero value for \bG,
no matter what the value of the strength c $\sim 1/U_0$ is,
a straight zero horizontal axis line $\gamma(\Gamma^+ = 0,0 \leq c \leq 1)$ exist;
where the imaginary part disappears and the scattering phenomenon losses any physical meaning.
According to our interpretation, it doesn't imply that the quasiparticles lifetime $\tau(0)$ is infinite,
just it´s physically nonentity.

As soon as small finite values of the impurities concentration appears ($\Gamma^+ \sim n_{imp}$),
the 3D manifold emerges (Fig.\ref{3}) with the maximum value for the $\gamma(\Gamma^+,c)$ surface in the
neighborhood of the point with $\gamma(\Gamma^+=1,c=0)$ and a gap value $\Delta_0=0$, $\gamma_i = 1$,
and $\gamma(\Gamma^+=1,c=0) = \pi$.
At this point, we find the smallest lifetime $\tau(0)$ in the
unitary regime and $\omega=0$ (a red colored neighborhood on
the right side in Fig.~\ref{3}). But for values around the
point $\gamma(\Gamma^+=1,c=1)=1.57$ smaller $\gamma$ means that
a bigger lifetime $\tau(0)$ exist with $\omega= 0$,
corresponding to a Born scattering with some remaining
ill-quasiparticles states (green colored neighborhood on the
left side of Fig.~\ref{3}).
Another case of interest is when $\tilde{\omega} \neq 0$ has a
constant value and the 3D manifold $\gamma(0 \leq \Gamma^+ \leq
1,c=0)$ converts into a 2D straight line $\tilde{\omega} =
\omega -i \; \gamma(\Gamma^+,c=0)$. In this case $\omega \neq
0$ has a very large potential (and a greater slope), with \bG
equivalent to the Landau Lifshitz parameter $\Gamma$ mentioned
in their Quantum Mechanical Book, Chapter XVII, section 132,
equation 132.1 \protect\cite{19} which resembles the unitary
limit. Finally in Fig.~\ref{3} there is another case at
$\tilde{\omega} = \omega - i \;\gamma(\Gamma^+,c=1)$, with a
straight line of smaller slope, corresponding to a Born limit
with $\omega \neq 0$ and a very small potential
\protect\cite{19}. The case when $\Delta \neq 0$ will we
analyzed in a future work, here our discussion follows a zero
temperature approach.

Following the analysis we calculate equation (2) corresponding to the self-frequency consistent \bw using the minimization algorithm.
This equation contains the inverse of the average lifetime states for the quasiparticles bound to the impurity \protect\cite{5,12,hir1}.
As Fig.~\ref{4}(a)-top frame and Fig.~\ref{4}(b)-bottom frame show,
two physical limits are well established from the imaginary part of \bw: the Born and the unitary regimes.

In both cases, the figures display the unitary regime with the strength parameter interaction at $c = 0$
in black color. In such a case, the imaginary part of the frequency (the disintegration probability per unit time) $Im [\tilde{\omega} (\omega)]$
has a maximum at zero frequency $\omega = 0$. This physically shows that the magnitude of the scatter potential U$_0$
is strong enough to immediately break the bound state of the Cooper pair since $c = 1 / (\pi \; N_F \; U_0)$.

Moreover, as the interaction parameter c moves away from the unitary limit (with $\tau(0) \rightarrow 0$),
the maximum of the disintegration probability $Im [\tilde{\omega} (\omega)]$ decreases as the parameter c
increases (the interaction potential U$_0$ decreases) as a function of the frequency $\omega$.

From values for $c \geq 0.2$, it´s observed that the behavior is close to the Born limit,
for which the value of the interaction potential U$_0$ is small in comparison with the electronic energy; therefore $\tau(\omega)$ increases.
Top frame of Fig.~\ref{4} should be compared with Fig. 1 bottom frame in reference \protect\cite{12},
while the bottom frame in Fig.~\ref{4} is showed and discussed here for the first time.

The bottom part of Fig.~\ref{4} shows that with very small \bG $= 0.001 meV$,
the disintegration probability $Im [\tilde{\omega} (\omega)]$ is very small for most of the whole frequency range and
the maximum $Im [\tilde{\omega} (\omega)]$ shifts with a pronounced peak to the right values of the frequency $\omega$.

\newpage

\section{Superconducting density of states}\label{sec:dos}

In this section, the DOS results are presented. In unconventional superconductors,
the order parameter goes to zero at some parts of the Fermi surface. Due to this fact, the density of states at very low energy
arises from the vicinity where the nodes of the order parameters are located.
Well known examples of this are the (HTS) \protect\cite{8}.
In general, line nodes and point nodes give a density of states that varies at the low energy limit as $\omega$ and $\omega^2$ respectively
\protect\cite{5,p1}.

Besides the nodes in the order parameter, scattering from non-magnetic impurities also influences the calculation of the low energy DOS
\protect\cite{5,hir1,11}. This elastic scattering mechanism leads to the lowering of T$_c$; and therefore to the suppression of the superconducting state.
In general, for temperatures much smaller than T$_c$, the effect of having very low nonmagnetic impurities concentration can be neglected.
It is found that only for very low temperatures, the effect of impurities becomes important for the unitary limit.
However, for clean samples, this effect can be neglected \protect\cite{hir1,p0}.

Therefore in dirty d-wave (HTS) becomes interesting to reconstruct the appearance of normal states at zero temperature.
Hence we study the evolution of the gap as a function of \bG. (we used the minimization routine \protect\cite{14,17}).
The DOS for low frequencies ($\omega \rightarrow 0$) is calculated by the following expression \protect\cite{5}:

\begin{equation}
N (\omega) = N_{FS} \; \re [ \; g(\tilde{\omega} (\omega)) \; ].
\label{ds1}
\end{equation}
	 		
Where $g (\tilde{\omega} (\omega))$ is given by (3), the Fermi average $<...>_{FS}$ on equation (3) is performed over a spherical Fermi surface with
a polar representation of the d-wave gap $\Delta = \Delta_0 \cos (2 \theta )$, see Fig.~\ref{1}.
Experimentally, the value for  $\Delta_0 = 24 \sqrt{2} (meV)$ corresponds to the maximum
value obtained by the $ARPES$ technique \protect\cite{12,13}.

\begin{figure}[h]
\begin{center}
\includegraphics[width = 4.5 in, height= 4.5 in]{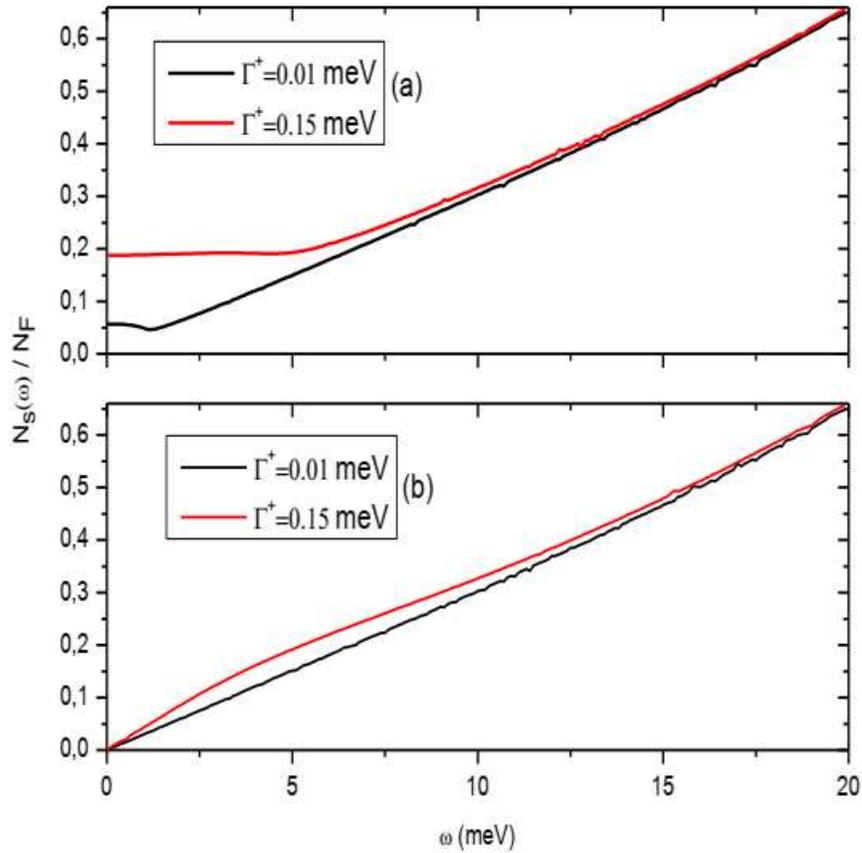}
\end{center}
\caption{\label{5} The DOS normalized $N(\omega) / N_F$ for top bottom-(a) unitary regime. and low bottom-(b) Born case}
\end{figure}

As the fourth result, Fig.~\ref{5}(a) top frame shows the unitary limit case with an a $c = 0.0$ strength parameter
(where a single Cooper pair scatters many impurities, and $n_{imp}$ is larger (\bg $= 0.15 meV$);
it remarkably shows that there is a large density of residual states $N(0)$ for higher concentrations of impurities
\bG $\sim n_{imp}$ (red line) mapping the zero frequency $\omega = 0$ DOS unitary behavior.
However as this concentration decreases an order of magnitude, the amount of residual states $N (0)$ also decreases Fig.~\ref{5}(a, black line).
Regardless of this, the same behavior persists for both curves, represented by the black and red colors
corresponding to the unitary regime.
The residual $N (0)$ existence changes the behavior in the universal limit of certain kinetic coefficients
such as the thermal conductivity $\kappa_ij (T)$ and the sound attenuation $\alpha_ij (T)$ \protect\cite{5,p0,p2}.

Bottom frame of Fig.~\ref{5}(b) corresponds to the case with an intermediate scatter parameter
$c = 0.2$ which represents in practice the dispersion due to a single impurity, and
the bound of the U$_0$ potential with the Cooper pair is relatively small (Born limit).
We see from Fig.~\ref{5}(b) that there is no residual density of states $N(0)$ at zero frequency
at it was shown in ref. \protect\cite{12} for HTC d-wave superconductors,
and in ref.\protect\cite{hir1} for heavy fermion superconductors.
For the calculation of the DOS in this work, we use two \bG values to test the non-linear minimization technique.

\newpage

\section{Conclusions}\label{sec:conclu}

This study compares and implements two algorithms to solve the imaginary non-linear
self-consistent equation for \bw in non-magnetic unconventional superconductors.
As a main conclusion, this work states that the minimization algorithm \protect\cite{14,17}
is the appropriated numerical procedure to solve \bw in the complex field,
when the number of varying parameters are two, that is to say, the impurity concentration \bG and the
strength parameter c. The cost in time showed in Table 1 correctly approaches the space parameter minimization
search accounting for different \bG values. On the other hand, the self-consistent fixed-point algorithm \protect\cite{16} only manages to solve
well equation (2) for c parameter (faster according to Table \ref{table1}, but as \bG changes it does not converge). That means that the fixed-point method is not suitable for a parameter
space of dimension $d \geq 2$. This result is based on the exact reproduction of the imaginary part of \bw, as well as,
the density of superconducting states DOS previously calculated in ref. \protect\cite{12} for different values of \bG.

Additionally, it was possible to obtain the inverse of the average lifetime $\tau(\omega)$
and the imaginary part (disintegration probability) $Im [\tilde{\omega} (\omega)]$ in Fig.~\ref{4}-(b)
for very small impurity concentration values of \bG such as \bG $= 0.01 meV$.
This result offers a relevant approach to numerically study the unitary limit regime.
To a certain extend, there are strong experimental evidences that the heavy fermions superconductors
UPt$_3$ \protect\cite{10,hir1} and UPd$_2$Al$_3$ \protect\cite{ult} are in the unitary regime.
This calculation is not reported in reference \protect\cite{12} (the authors did not mention
their numerical approach to solve equation (2)). Furthermore, this report points out the importance of the use of
appropriate numerical methods such as the minimization procedure \protect\cite{14,17}
when studying elastic scattering quantum mechanical phenomena \protect\cite{19,ult2}
with parameter space of dimension $d \geq 2$.

Finally, this study emphasizes that the physics of the unitary regime in superconducting metals
needs further clarification \protect\cite{ult2,ult3} as we stated in section~\ref{sec:nr}
where Fig.~\ref{3} is analyzed from a more general quantum mechanical approach
(the footnote is added in the bibliography) \cite{19,ult,ult2}.

\newpage

\section*{Acknowledgments}

One of the authors, P. Contreras wishes to express his gratitude to Professor Kirill Samokhin from Brock University, who enlightened him
the theoretical subject of this paper several years ago.
The authors also acknowledge Dr. David Luxat for several discussions regarding the numerical part of this work, and
Prof. Andre-Marie Tremblay from Universite de Sherbrooke who kindly suggested some relevant literature.


\end{document}